\documentclass[twocolumn,prb,showpacs]{revtex4}
\usepackage{epsfig}
\usepackage{dcolumn}
\usepackage{bm}
\newlength{\defaultparindent}
\newenvironment{Body Text 2}{}{}
\newenvironment{Body Text}{}{}
\newenvironment{Default Paragraph Font}{}{}

\begin{document}
\title{Magnetic  polarons in weakly doped high-$T_c$
superconductors.}

\author{E. M. Hankiewicz, R. Buczko, and Z. Wilamowski}

\affiliation{Institute of Physics, Polish Academy of Sciences, al.
Lotnik\'ow 32/46, PL 02--668 Warsaw Poland}

\begin{abstract} We consider a spin Hamiltonian describing $d$-$d$ exchange
interactions between localized spins $d$ of a finite antiferromagnet as well as
$p$-$d$ interactions between a conducting hole ($p$) and localized spins. The
spin Hamiltonian is solved numerically with use of Lanczos method of
diagonalization. We conclude that $p$-$d$ exchange interaction leads to
localization of magnetic polarons. Quantum fluctuations of the antiferromagnet
strengthen this effect and make the formation of polarons localized in one site
possible even for weak $p$-$d$ coupling. Total energy calculations, including
the kinetic energy, do not change essentially the phase diagram of magnetic
polarons formation. For parameters reasonable for high-$T_c$ superconductors
either a polaron localized on one lattice cell or a small ferron can form. For
reasonable values of the dielectric function and $p$-$d$ coupling, the
contributions of magnetic and phonon terms in the formation of a polaron in
weakly doped high-$T_c$ materials are comparable. \end{abstract}

 \maketitle


\section{INTRODUCTION}
The role of electron phonon and spin exchange in the formation of
polarons, bi-polarons and stripes in CuO$_2$ based high-$T_c$
materials has been a matter of extensive discussion from the very
moment high-$T_c$ superconductors were discovered. Experimental
observations of the anomalous isotope
effect\cite{Crawford90,Franck91} could indicate the role of phonon
interactions in the formation of the superconducting state. On the
other hand, angularly resolved photoemission \cite{Norman00}
(ARPES), transport \cite{Karpinska00} and tunneling microscopy
measurements \cite{Barbara00} show $d$-symmetry of the order
parameter indicating an important role of exchange. Although some
authors underline the role of both mechanisms \cite{Castellani95},
the solution of full Hamiltonian  is complicated and  these
contributions are usually calculated separately.

The concept of the phonon polaron was introduced by Pekar \cite{Pekar51} and
Fr\"ohlich \cite{Froilich54a,Froilich54b} decades ago and was adopted to
high-$T_c$ superconductors by Alexandrov and Mott \cite{Alexandrov95}. The
energy of phonon coupling for high-$T_{c}$ is estimated  to be of the order of
a fraction of eV. In another work, Alexandrov \cite{Alexandrov88} proposed that
the narrowing of $d$ -electron band by the polaron effect and the formation of
phonon bipolarons increases the critical temperature to $T_c\approx100$~K.
However, because of the dominant contribution of $d$-symmetry to the order
parameter of high-$T_c$ superconductors, exchange interactions are the most
often considered ones recently. The modeling of such interactions is difficult
because it is necessary to include the role of conducting $(p)$ holes as well
as localized $(d)$ electrons.

In the case of weak $p$-$d$ couplings a linear response of the
antiferromagnet (AF) can be assumed. Hence, a phenomenological
model of magnetic susceptibility introduced by Millis, Monien,
Pines \cite{Millis91} (MMP) can be used.

 Zhang and Rice \cite{Zhang88}, in turn, describe the case of
strong hybridization between the $3d$ states of Cu and $2p$ states of O. When
the hybridization is much greater than the kinetic energy of the carriers, it is
possible to limit the description to the ground singlet state, and  in such a
case a one band model can be used. This model (called $t$-$J$) is at present
the most often used to describe the cuprate superconductors (for a review see
\cite{Dagotto94}). Unfortunately, different techniques of approximations lead
sometimes to contradictory solutions of $t$-$J$, as it is in the case of the
stripe formation \cite{White00,Hellberg00}. Also, the dynamics of spin polaron
in the framework of $t$-$J$ model was extensively investigated by many
authors\cite{Rink88,Kane89,Elser90,Ramsak93,Reiter94,Liu92}.

However, the spin-fermion model
\cite{Buhler00,Altshuler95,HankiewiczPB,Schrieffer95,Ramsak90,Shimahara95,Bala96,Bala00},
which treats separately the spin of $p$-carrier and $d$-localized spins, seems
to be the most appropriate for the case of intermediate $p$-$d$ coupling
typical for high-T$_c$ superconductors. Most  papers treat the $p$-$d$
interaction  as a Kondo exchange. Such an approach allows to show the tendency of
cuprates to stripe  phase formation \cite{Buhler00} or d-wave pairing
\cite{Schrieffer95, HankiewiczPB}. Recently, Ba{\l}a et al. \cite{Bala96,Bala00}
extended the spin-fermion model starting from the three band Hubbard model. They
introduced $p$-$d$ coupling, which is much more complicated than Kondo one, i.e.
some exchange couplings between Cu ion and two neighboring oxygen atoms appear.
They have considered the formation of magnetic polaron in such a
model\cite{Bala96,Bala00}.  They identified  all the spectrum of energy states
of Zhang-Rice polaron \cite{Bala00} in  ARPUS experiments \cite{Sawatzky97}.
They found a solution using self-consistent Born approximation (SCBA). However, a
comparison of SCBA and exact diagonalization techniques for spin-fermion model
has not been carried out yet.

The present paper reports for the first time  on   calculations of
a simple spin-fermion model with   exact diagonalization technique
for spin Hamiltonian. It broadens the possible range of solutions.
The spin-fermion model allows to investigate the cases of strong
and intermediate $p$-$d$ coupling. Thus a comparison of the phonon
and magnetic contributions to the formation of a polaron in weakly
doped high-$T_c$ superconductors for a wide range of parameters is
possible. The  magnetic exchange interactions are calculated
strictly while the phonon contribution is evaluated in the
framework of the Fr\"{o}hlich theory
\cite{Froilich54a,Froilich54b,Alexandrov95}.

In Section~2, we describe the Hamiltonian and the method of finding  the ground
and excited states of the system. In Section~3,  the exchange spin interactions
are analyzed. In Section~4, we construct and analyze phase diagram for the
formation of different magnetic polarons. In Section~5, we compare the energies
of phonon and magnetic polarons. A summary is given  in Section~6.

\section{MODEL}

We study a system which consists of spins localized on the $d$-shells and
mobile $p$-like holes. The localized spins are treated within the quantum
antiferromagnet model, i.e.  none of the Neel sublattices are chosen a
priori.  We consider  finite one- (1D) and two- dimensional (2D) clusters as
well as the infinite system. The latter is simulated  using periodic boundary
conditions (PBC) and an extrapolation to an infinite size. The spin
interactions in the system and the kinetic energy of the $p$ hole is described
by the Hamiltonian:

\begin{eqnarray}\label{e1}
 H=  2J_{dd} \sum\limits_{<i,j>} \mathbf{S}_{i} \mathbf{S}_{j}
   +\frac{1}{2}J_{pd} \sum\limits_{i,\alpha,\beta } \mathbf{S}_{i}
    c^{+}_{i\alpha } \bm{\sigma}_{\alpha \beta} c_{i\beta }
+
 \nonumber \\
+t_{ij} \sum\limits_{i,j,\alpha }c^{+} _{i\alpha } c_{j\alpha },
\end{eqnarray}
where
 $\bf{S}_{i} $
 is the effective spin of the nine core $d$-electrons (3d$^{9}$), the summation was taken under
 the pairs of neighboring spins $<i,j>$, $\bm{\sigma} _{\alpha \beta } $
are the Pauli matrices of the carrier spin $\mathbf{s}$, $J_{pd}$ is the Kondo
parameter normalized to the volume of elementary cell,
 $J_{dd}$ is the
exchange integral between $d$-$d$ spins, $t_{ij}$ is the hopping
matrix element to the neighbor,
 $c_{i\alpha } ^{+}$, $c_{i\alpha } $ are the operators of creation
 and annihilation of a hole on the $i$-th site, with the projection
 of spin $1/2$ on the quantization axis equal
$\alpha$ respectively.
 We look for the
eigenstates of the Hamiltonian (1) in the form of the product of
the orbital wave vector, $\Psi$, and  the  spin state vector
 of local and carrier spins $X_{S}$:

\begin{equation}\label{e2}
 \Psi \cdot X_{S}.
\end{equation}
In the space of the single hole occupation at antiferromagnet (AF) sites $i =1,
..., N$
 \begin{equation}\label{e3}
 \Psi =\sum\limits_{i}\varphi \left( i\right)u_i,
\end{equation}
where
 $u_i = \left| 0,0 ,...,1,0,0...\right\rangle$
is the occupation vector of the hole at $i$-th site and $\varphi(i)$ is the
corresponding probability amplitude. We use the variational method of solving
the eigenproblem. We treat $\Psi$ as the trial function, which  is normalized by
the condition:
\begin{equation}\label{e4}
 \sum\limits_{i}^N\left| \varphi (i)\right| ^{2} =1.
\end{equation}
For $\Psi$ given by Eq.~(2) the diagonalization of the kinetic and exchange
terms of Hamiltonian (1) can be done separately.
\begin{equation}\label{e6}
 H^\prime =\left\langle \Psi \right| H\left| \Psi \right\rangle =T+H_{s}.
\end{equation}
The kinetic energy, $T$, does not depend on the spin state $X_S$
and can be easily evaluated from the probability amplitude of the
carrier
 $\varphi \left( i\right)$:
 \begin{equation}\label{e8}
T=z\left| t_{1} \right| -2\left| t_{1} \right|
\sum\limits_{<i,j>}\varphi (i)\varphi (j),
\end{equation}
if only the hopping to the nearest neighbors (NN), $t_{1}$ is considered, and
$z$ is the number of NN. The term $z|t_{1}|$ has been added to have zero
kinetic energy for fully delocalized states. Calculations of the exchange
energy need diagonalization of the spin Hamiltonian:
\begin{equation}\label{e9}
 H_{s} = 2J_{dd} \sum\limits_{<i,j>}\mathbf{S}_{i} \mathbf{S}_{j}  +\sum\limits_{i}j_{pd}
(i)\mathbf{S}_{i} \mathbf{s}.
\end{equation}
Here $\bf{s}$ is the spin operator of the carrier spin and the $p$-$d$ exchange
integrals are given by
\begin{equation}\label{e10}
 j_{pd} (i)=J_{pd} \left| \varphi (i)\right| ^{2}.
\end{equation}

We use Lanczos method of  diagonalization of the spin Hamiltonian (7).
Diagonalization of matrices for AF cluster with $N \leq 20$ is possible with
the PC computer. For larger clusters the calculation time  increases rapidly.

A detailed study of quantum coupling between spins within the polaron in an AF
medium is the main goal of the paper. We focus our analysis on the gain
$E_{ex}$ of the spin exchange energy.
 $E_{ex}$  is the difference
between the energy of the ground state of the unperturbed AF $E_{dd}^{0}$ (when
$J_{pd}=0 $) and the exchange energy of the whole system when the carrier is
coupled to  AF $E_{dd}+ E_{pd}$, i.e.
\begin{equation}\label{e11}
E_{ex}  = E_{dd}^0 - (E_{dd}+ E_{pd}).
 \end{equation}
In this way, we analyze the zero temperature case only. We found that for a weak
$p$-$d$ exchange  the gain  $E_{ex}$ is a half of the $p$-$d$ coupling energy,
$E_{ex} =\,-E_{pd} /2$.

The Hamiltonian  includes neither effects of external magnetic
field nor any other terms which could lead to the symmetry
breaking and to the formation of Neel sublattices. As a
consequence, the mean values of the magnetic moments of each spin
of this  system are exactly equal zero. Because of this, we
describe the spin structure of the antiferromagnetic polarons
(AFP) and the mechanism of the magnetic polaron formation  by
means of correlators between spins of AF $\langle \mathbf{S}_i
\mathbf{S}_j \rangle$ and between the AF spin and that of an
additional hole $\langle \mathbf{s} \mathbf{S}_{i} \rangle $:
\begin{equation}\label{e13}
\langle \mathbf{s} \mathbf{S}_{i} \rangle = \langle
X_{S}|\mathbf{s} \mathbf{S}_i| X_{S} \rangle.
 \end{equation}

The assumption that the wave  function can be described by Eq.~2 is simplified.
In general, the total wave function is a linear combination of wave functions
given by Eq.~2. In particular, because polarons localized on two Neel
sublattices are equivalent, at least the combination of the pair of such states
should be considered. This problem is discussed in section IIIG.

Hamiltonian (1) does not contain a phonon term. In general, the coupling of
the carrier to the lattice polarization and the exchange interactions act
together to form the polaron.
 The phonon term  is treated in Section 5 within the simplest possible
approach. It is calculated from Fr\"ohlich interaction with
optical phonons. It is known \cite{Alexandrov95} that in the range
of parameters typical for  high-$T_c$ superconductors small phonon
polarons are formed. Their energy is given by:
\begin{equation}\label{e14}
 E_{ph} \approx \frac{q_{D} e^{2} }{\pi \kappa },
 \end{equation}
where
 $q_{D} =\left( \frac{6\pi ^{2} }{V} \right) ^{1/3} $ is the Debye momentum,
 $\frac{1}{\kappa } =\frac{1}{\varepsilon } -\frac{1}{\varepsilon_0} $;
  $\varepsilon$, $\varepsilon_0$ are  dynamic and  static dielectric constants respectively,
  and $e$ is the electron charge.
\begin{figure}[tbh]
\centerline{\epsfig{file=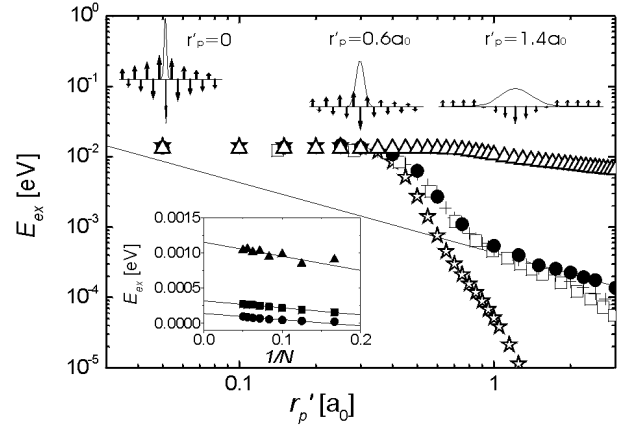, width=8.cm,clip=}}
\caption[]{The
 exchange energy gain, $E_{ex}$, obtained for a
Gaussian trial function  is presented as a function of polaron
radius
 $r'_p$, for 1D open chain (crosses), for 1D chain with (PBC) (open squares) and for a $4\times4$
square with PBC (open stars). Open triangles show $E_{ex}(r'_p)$
for a comb-like carrier trial function (see Eq. \ref{e22b}) and 1D
AF chain of 16 spins. In the inset the extrapolation of $E_{ex}$
to infinite AF 1D cluster is shown by filled triangles, squares
and circles for $r'_p =0.75a_0$, $r'_p=1.5a_0$ and $r'_p =3a_0$
respectively, where $a_0$ is a lattice constant. In the icons
arrows show  the correlators between the carrier spin and  spins
of AF for a small AFP ($r'_p = 0$), medium AFP ($r'_p =0.6a_0$)
and large AFP ($r'_p =1.4a_0$). For a large AFP, the arrow size is
multiplied 30 times for clarity. Solid curves show trial
functions.}
\end{figure}

\section{SPIN STRUCTURE OF MAGNETIC POLARONS}

We examine various types of trial functions $\varphi$. Depending on the size
and shape of the carrier distribution, $|\varphi|^2$ (Fig.~1), as well as on
the value of exchange constants (Fig.~2) various types of magnetic polarons
with very different spin structures can be found.

\begin{figure}[t]
\mbox{}\centerline{\epsfig{file=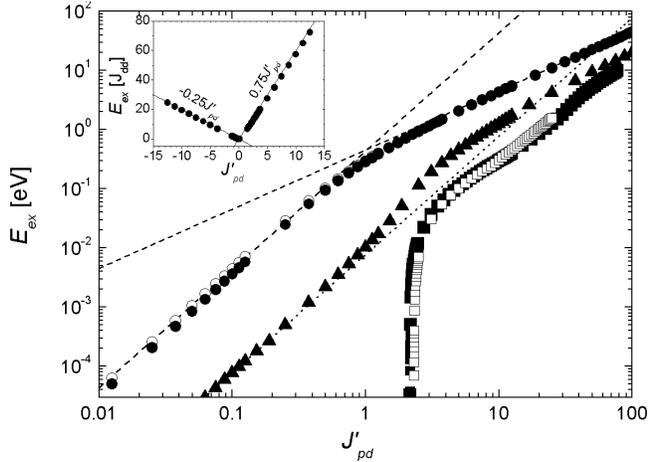,width=8.5cm}} \caption[]{The exchange
energy gain, $E_{ex}$, from formation of different magnetic polarons  as a
function of $J'_{pd}$. $E_{ex}$ for a small AFP were obtained  for a 1D chain
with PBC (open circles) and for a 2D $4\times4$ square with PBC (filled
circles). $E_{ex}$  for a  trial function evenly distributed on all AF spins
are shown by open and filled squares for a 1D chain with PBC and 2D $4\times4$
with PBC respectively. Results for a trial function distributed on two neighbor
AF spins and a 1D chain with PBC are drawn by filled triangles. The dashed
lines show  quadratic and linear dependencies of $E_{ex}$ on $J'_{pd}$. The
dotted line emphasizes superquadratic dependence of $E_{ex}(J'_{pd})$.
 The inset shows  $E_{ex}(J'_{pd})$ for a small AFP.}
\end{figure}

{\bf A.\ Large AFP}. Its size $r'_p$  is much greater than the
lattice constant $a_0$ and the antiferromagnetic correlation range
$\xi$. Thus, it can be described within the theory of the linear
response. For very large polarons the local magnetization
$M(\mathbf{r})$ becomes proportional to the local carrier density
\cite{Millis91}.

 {\bf B.\ Medium AFP}. Its size $r'_p$ is comparable or smaller
than $\xi$, but bigger than $0.5a_{0}$ (see Fig.~1). The induced
polarization is of staggered character, and the carrier interacts
with few local spins of different orientation.

 {\bf C.\ Small AFP} It is  localized within a
single elementary cell. For very strong $p$-$d$ coupling it
corresponds to the Zhang-Rice polaron.

 {\bf D.\ Ferron}.
It forms when the $p$-$d$ exchange is strong enough to break AF
bonds and to induce a local magnetic moment \cite{Kasuya68}.

 {\bf E.\ Comb like AFP} It is the  case when the carrier is distributed
on one Neel sublattice only. The staggered polarization is the
dominant response of the AF medium
\cite{HankiewiczPB,HankiewiczACT}.

The clear difference between AFP of different sizes is seen in Fig.~1, where
the energy gain, $E_{ex}$, is plotted as a function of the polaron size
$r'_{p}$. Gaussian trial function $ |\varphi \left( i\right) |^{2} = A\exp
\left[ -\frac{\left( \bf{r}_{i}-\bf{r}_{0} \right) ^{2} }{2r_{p}^{'2} }
\right]$, where $A$ is a normalization factor and $\mathbf{r}_{i}-\mathbf{r}_0$
 is the distance from central spin, was used to calculate  almost all curves.
  The curve plotted by triangles, however,
 was calculated using another trial function, and it  represents
 the formation of a comb-like AFP. It will be discussed in part III~E.

The calculations were performed for weak $p$-$d$ coupling for 1D and 2D AF
clusters.  The number of AF bonds has been assumed to be twice as big for 2D as
for a 1D cluster, where the number of ligands is two times smaller.  To compare
the results for 1D and 2D clusters we introduce the following quantity,
$J'_{pd}=J_{pd}/(2z J_{dd})$, which describes the ratio of $p$-$d$ to $d$-$d$
coupling normalized by the number of AF bonds, $2z$. Ratio of the $p$-$d$ to
the $d$-$d$ exchange energies defined this way is the same for 1D and 2D AF
clusters. Moreover, $J'_{pd}\ll1$ describes the range of weak $p$-$d$ coupling,
while $J'_{pd}>1$ that of strong $p$-$d$ coupling, independent of the dimension
of the AF cluster.  $J'_{pd}$ is set 1/4 in Fig.~1. Here the small AFP
corresponds to $r'_p <0.4 a_o$, medium to $0.4 a_o<r'_p<a_o$, and large to
$r'_p > a_o$. The dependence of $E_{ex}$  on the chain length is shown in the
inset. The linear dependence of the $E_{ex}$ on the inverse AF cluster size
$1/N$ makes the extrapolation to an infinite chain possible. The extrapolated
values of $E_{ex}$ are shown by full circles in Fig.~1.

The results of calculations  for  small and medium AFP in the 1D case are
reasonable  even without extrapolation to infinite clusters. Because of that, we
assume that also for the 2D case the energy of small and medium polarons can be
found with reasonable accuracy. Unfortunately, because of computational
limitations, we can not carry out full extrapolation for large AFP in 2D
clusters.

A comparison of numerical results for open chains and chains with
PBC shows that the use of PBC does not lead to an improvement of
the convergence of numerical results with the cluster size. In
part, this is an effect of the normalization condition given by
(Eq.~4) and of specific properties of the dangling spins at the
cluster border.

\subsection{Large AFP}

The numerical results, with the characteristic dependence of the $E_{ex}$
decreasing as $1/r'_p$ for 1D AF, can be approximated by an analytical solution
in which the magnetic susceptibility of AF is described by two phenomenological
parameters. The analytical solution gives a better description of large AFP
while the comparison with the numerical results allows to explain the physical
meaning of the phenomenological parameters.

For simplicity in the analytical approximation, we use an
exponential distribution of carrier density,
$\mathbf{|\varphi(r')|}^2=\frac{1}{2r_p}\exp(-|\mathbf{r'}-\mathbf{r}_{0}|)/r_p$,
where $r_0$ is the center of the density distribution and
$r_p^2=2{r_p^2}'$.
 We assume that
the susceptibility in wave vector space can be described by the
phenomenological formula introduced in the MMP model
\cite{Millis91},
$\chi(\mathbf{q})=\chi_{q_0}\frac{1}{1+(\mathbf{q}-\mathbf{q}_0)^2\xi^2}$,
where $\mathbf{q}$ is the wave vector, $q_0 =\pi /a_0$ and
corresponds to the staggered magnetization. The phenomenological
parameters $\chi_{q_0}$ and $\xi$ describe the staggered magnetic
susceptibility and the spin correlation length, respectively. An
effective field corresponding to $\varphi(\mathbf{r'})$ is
$H_{eff}(\mathbf{r'})=\frac{J_{pd}a_{0}}{4g\mu_Br_p}\exp(-|\mathbf{r'}-\mathbf{r}_0|/r_p)$.
Within the model of continuous media, an effective field acting at
a point $\mathbf{r'}$
 results in magnetization $M(\mathbf{r})=\int dr'\chi(\mathbf{r-r'})H_{eff}(\mathbf{r'})$
 at the point $\mathbf{r}$. Such an approach takes into account the non-local effects of
correlated systems but neglects the atomic structure of AF. For
large AFP, i.e.\ for $r_p\gg\xi$,  $M(\mathbf{r})$ takes a simple
form:
\begin{equation}\label{e18}
  M(\mathbf{r})=\frac{J_{pd}a_{0}\chi_{q_0}}
  {16\pi^2
  g\mu_B\xi^2}\frac{1}{q_0^2r_p}\exp\left(-\frac{|\mathbf{r}-\mathbf{r_0}|}{r_p}\right).
\end{equation}
Thus $M(\mathbf{r})$ induced by a large AFP is smooth and
described by the same spatial dependence as the carrier density.
In the icon in Fig.~1, the correlators between the carrier spin,
$\mathbf{s}$ and the consecutive local spins, $\mathbf{S}_i$, for
a large AFP are presented. They also show that for a large AFP the
induced magnetization is smooth and proportional to the carrier
density. In that sense the analytical solution is equivalent to
the numerical one.  $E_{ex}=1/2E_{pd}$ for a large AFP and it
takes a simple form:
\begin{equation}\label{e19}
E_{ex} = \frac{(J_{pd}a_{0})^{2} \chi _{q_0} }{256\pi^2\left( g\mu
_{B} \right) ^{2} \xi ^{2} q_{0}^{2} r_{p} }.
 \end{equation}

The absolute value of  $E_{ex}$ and its dependence on the $d$-$d$ exchange and
on the correlation length cannot be analyzed within the discussed
phenomenological approach. However, an evaluation of $E_{ex}$
 can be done by numerical calculations. If one attributes the correlation
length to  the cluster size, then the comparison of  Eq.~\ref{e19} with the
numerical result, showing that the energy gain  is almost independent of the
chain length (see inset to Fig.~1), leads to the conclusion that $\chi_{q_0}
\propto\xi^{2}/J_{dd}$ for  a large 1D polaron.
 We found out that the character of the obtained dependencies
 does not change if a Gaussian trial function was applied.

In the 2D case, the results of numerical
integration for large AFP in a 2D system are characterized by a
faster decrease of  $E_{ex}$ with the polaron size in comparison to the 1D
case. The energy scales with $1/r_p^2$ instead of $1/r_p$.

To summarize, the large AFP is characterized by a smooth
magnetization of the antiferromagnet, proportional to the carrier
density. Because for a 2D polaron
 $E_{ex}$ and the kinetic energies scale with
the inverse square of the polaron size, two scenarios are possible: (i) no
polaron can be formed when $T>E_{ex}$, or, in the opposite case, (ii) the
minimum of the total energy occurs for the polaron radius tending to zero (see
also Fig.~6). In that case, however, it is no longer a large polaron and a
different approximation should be considered.

\subsection{Medium AFP}

As it is shown in Fig.~1, with a decrease of $r'_p$, $E_{ex}$ does not follow
the $1/r'_p$ dependence, but for $r'_p<a_o$ a sharp increase of $E_{ex}$ is
seen. For 1D the increase of $E_{ex}\sim10$, as compared to the extrapolated
$1/r'_p$ dependence, and for 2D  is even bigger.The rapid increase of $E_{ex}$
shows that the medium AFP has a tendency to localization. The decrease of $r'_p$  is
also accompanied by a change of the character of induced magnetization in the
AF medium. As it is shown by icons in Fig.~1, numerical calculations predict
 a smooth magnetization for $r'_p \geq a_0$  but a staggered one for
 $r'_p < a_0$.

For a weak $p$-$d$ coupling,  $E_{ex}$
 increases with the square of
$J'_{pd}$ (see filled triangles in  Fig.~2).  However, calculations show that
for
 $J'_{pd}>1$ a more complex mechanism occurs. The AF bonds  broke
and a ferron-like polaron forms. The details are described in part
IIID.

\subsection{Small AFP}

When the polaron radius becomes comparable to the size of the elementary cell,
 $E_{ex}$ saturates. The plateau in  Fig.~1
corresponds to the case when the polaron is localized in a single elementary
cell, i.e.
 $r'_p<a_{0} $. The numerical results weakly depend on the size of AF
  cluster and on boundary conditions. The magnetization induced by the small AFP is similar to that of the medium
AFP, as it is shown by icons in Fig.~1.

\begin{figure}[t]
\centerline{\epsfig{file=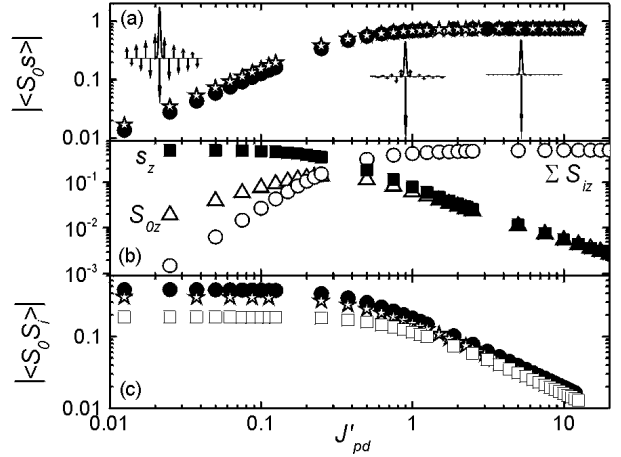, width=8.cm,clip=}}
\caption[]{(a) The correlator,
$|\langle\mathbf{s}\mathbf{S}_0\rangle|$, between the carrier
spin, and AF spin on which the carrier is localized as a function
of $J'_{pd}$ for 1D chain with PBC (filled circles) and 2D
$4\times4$ square with PBC (open stars).  Icons show the
correlators between the carrier spin, $\mathbf{s}$, and
consecutive $d$ spins of AF chain for corresponding $J'_{pd}$. (b)
The z-component of carrier spin, $\mathbf{s}_z$, of central spin
$\mathbf{S}_{0z}$, as well as the sum of AF spins
 $\sum_{i}\mathbf{S}_{iz}$ are shown for an open chain of 16 spins as a function of
$J'_{pd}$. (c) The correlator between the spin on which the
polaron is acting and the NN (filled circles) and the NNN spin
(open squares) for a chain of 16 spins. The correlator between the
spin on which the polaron is acting and the NN spin for a 2D
$4\times4$ square with PBC is presented by stars.}\end{figure}

A dependence of the numerically calculated  $E_{ex}$
 on the normalized ratio $p$-$d$ to $d$-$d$ exchange
 $J'_{pd}$ is shown by circles in Fig.~2. There is no considerable difference
between 1D and 2D cases. For a weak $p$-$d$ coupling $E_{ex}\propto
{J'}_{pd}^2$. The upper limit of this quadratic dependence corresponds to the
case when the $p$-$d$ coupling is equal to the $d$-$d$ exchange. For higher
$J'_{pd}$
 the energy increases linearly.

The correlator $\langle \mathbf{s}\mathbf{S}_0 \rangle$ between the carrier
spin, $\mathbf{s}$, and AF spin  on which the carrier is localized,
$\mathbf{S}_0$, is shown in Fig.~3a. In the range of weak $p$-$d$ couplings, the
magnetization of the local spin, as seen by the carrier spin, $\langle
\mathbf{s}\mathbf{S}_0 \rangle$, increases linearly and saturates for the
strong $J'_{pd}$. This corresponds to quadratic and linear dependence of
$E_{ex}$ versus $J'_{pd}$,  respectively. For a strong $J'_{pd}$,
 the correlator $\langle \mathbf{s}\mathbf{S}_0 \rangle=-3/4$
 that means  AF  spin   is compensated by the spin of a carrier.

The icons in Fig.~3a show the correlators  between the carrier
spin $\mathbf{s}$ and the consecutive $d$ spins of AF chain for
different $J'_{pd}$. It is seen that a small polaron induces a
staggered magnetization within the AF correlation range, $\xi$.
But the amplitude of $\langle \mathbf{s}\mathbf{S}_0 \rangle$
varies with $J'_{pd}$. The increase of the correlator $|\langle
\mathbf{s}\mathbf{S}_0 \rangle|$
 is accompanied by a reduction of the correlators between the carrier spin and other d spins.
Also, the correlator of the spin
 $\mathbf{S}_{0} $
 with neighboring spins
 $\mathbf{S}_{i} $, $\langle \mathbf{S}_0\mathbf{S}_i \rangle$,
 decreases
 with $J'_{pd}$ increase. The dependence of correlator for NN and the next nearest neighbor (NNN)
  are shown in Fig.~3c. The correlators
 are not affected by the presence of a weak polaron, but for stronger $J'_{pd}$
 they decrease as
 $ 1/J'_{pd}  $.

The ground state of a system consisting of  a carrier with $s=1/2$ and an AF
cluster of an even number of  local spins  is a spin doublet,  for any
$J'_{pd}$.
 However, depending  on the value of $J'_{pd}$, the magnetic moment
 can be transferred from the carrier to the system of local spins.
  For a non-interacting spin $\mathbf{s}$, or for a very weak $p$-$d$ coupling, the
carrier spin has a  magnetic moment while the total  moment of localized spins
is zero. In Fig.~3b the $z$-component of the carrier spin, $s_z$, the local
spin
 $S_{0z} $
and
the sum of all local spins
 $\sum_{i}S_{iz}   $
 are plotted as a function of $J'_{pd}$.
 To distinguish the z-direction, we put a small magnetic field
which does not perturb the spin structure of polaron. With an
increase of $J'_{pd}$ the magnetic moment is transferred from the
carrier to the local spins. For a very strong $p$-$d$ coupling,
the  moment of local spins $\sum_{i}S_{iz}$ saturates at value
$1/2$. At the same time, the moment of carrier spin and of the
local spin,
 vanish.
The  magnetic moments of the local spins, $\sum_{i}S_{iz}$, are distributed in
a staggered way among the whole cluster. In the limit of strong $p$-$d$
coupling,
 $J'_{pd} \gg 1 $ the small polaron
 is equivalent to the Zhang-Rice (ZR) one.
 Here the spins, $\mathbf{s}$ and $\mathbf{S}_0$, are
compensated and form a local singlet state, not coupled to other
local spins, $\mathbf{S}_i$ (see Fig.~3c and the icon in
Fig.~3a.). On the other hand, the formation of a  central pair of
compensated spins is accompanied by a transfer of the magnetic
moment to the neighboring local spins and formation of staggered
$M(\mathbf{r})$ around the central cell. Such an effect was
described in details, for example in \cite{Laukamp98}. The
reconstruction of AF bonds in the vicinity of the small AFP is
driven by a gain of $d$-$d$ exchange energy. It is of the order of
$J_{dd}$. It is a small contribution to the whole energy of the ZR
polaron but becomes  important for
 $J'_{pd}\cong 1 $. Moreover,  the reconstruction of the AF environment should
be taken into account when  interactions between separate polarons
are discussed.

To summarize, the small AFP in the range of intermediate $p$-$d$
coupling (typical for high-$T_c$ superconductors) is not
equivalent to the ZR polaron. The carrier spin is also partially
correlated with AF spins in the vicinity of local singlet.

\subsection{Ferron}

        As it is shown in Fig.~1, $E_{ex}$
for a weak $p$-$d$ coupling decreases with an increase of the
polaron size. For big
 $J'_{pd} $, however, some new effects occur. As it is shown
in Fig.~2, for a weak polaron evenly localized on two $d$-spins
antiferromagnetically correlated $E_{ex} $ is smaller by a factor $\sim80$ than
in the case of a small AFP. With an increase of
 $J'_{pd}$, instead of the expected saturation,
a faster increase of $E_{ex} $ for such a polaron
 occurs.

 An even more pronounced effect
is seen for a homogeneously distributed carrier density. As it is
shown by squares in Fig.~2,
 $E_{ex} $
 for a weak $p$-$d$ coupling is practically
zero but sharply increases for $J'_{pd}\simeq2$. Moreover, a further increase
of
 $J'_{pd}$ is accompanied with a step-like
increase of   $E_{ex} $. Studies of  the spin structure show that
this critical behavior is caused by a breakdown of the AF bonds
and the formation of a magnetic moment in the antiferromagnetic
medium. Such a magnetic polaron is known as ferron, and it was
described decades ago\cite{Kasuya68, Nagaev76}. In the whole range
of $J'_{pd}$ values there is no considerable difference between
the 1D and 2D cases.

        The spin structure of a ferron localized on two local
spins is complex. The correlator of the carrier spin $\mathbf{s}$
 with the two local AF spins $\mathbf{S}_p$,
  $\left\langle \mathbf{s}\mathbf{S}_{p} \right\rangle $,
is negative (for  antiferromagnetic $p$-$d$ coupling) for the
whole range of $J'_{pd}$. For a weak $J'_{pd}$,
  $\left\langle \mathbf{s}\mathbf{S}_{p} \right\rangle $
 increases linearly, then saturates for very large
$J'_{pd}$ approaching  $-0.5$. The sharper increase of $E_{ex}$, which is seen
for $J'_{pd}\cong 2.5\,\,$ is caused by a spin-flip of the two local spins,
$\mathbf{S}_{p} $. The systematic increase of  $E_{ex} $, faster than the
linear increase,  for $J'_{pd}
>4\,\,$
  originates neither from the correlation
 $\left\langle \mathbf{s}\mathbf{S}_{p} \right\rangle $
 nor from the correlation between two local spins
 $\left\langle \mathbf{S}_{p1}\mathbf{S}_{p2} \right\rangle $ but
 results from a systematic reconstruction of the AF
environment of the ferron. The correlation of each of the
 $\mathbf{S}_{p} $  with  surrounding
spins,  $\mathbf{S}_{i} $,  gradually decreases. Simultaneously,
there appears a reconstruction of $d$-$d$ couplings among the
surrounding spins.

\begin{figure}[bt]
\centerline{\epsfig{file=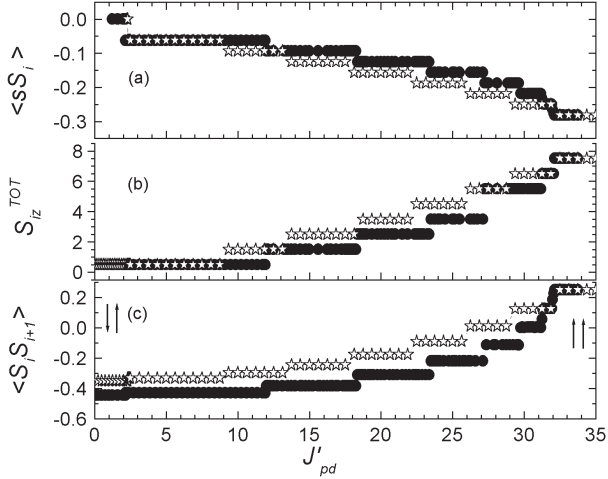, width=8.cm}} \caption[]{The
correlator between the carrier spin and one of AF spin $\langle
\mathbf{s}\mathbf{S}_i \rangle$ (a), the total spin moment of the
AFP polaron, $S_{iz}^{tot}$, (b), and the correlator between two
nearest AF spins $\langle \mathbf{S}_i\mathbf{S}_{i+1} \rangle$
(c) as a function of $J'_{pd}$ for a homogeneously distributed
trial function. The case of a 1D chain of 16 spins with PBC  is
presented by circles, the 2D $4\times4$ square with PBC is shown
by open stars.}
\end{figure}

         More apparent data about the mechanism of the formation of a ferron
can be found if one assumes a homogeneous distribution of the carrier on the
whole AF cluster with PBC. The data are shown in Fig. 4. The correlator between
one of the AF spins and the carrier spin as a function of $J'_{pd}$ is
presented in Fig.~4a. The step-like behavior of  $\langle
\mathbf{s}\mathbf{S}_{i} \rangle $ is visible for both 1D and 2D cases. For
$J'_{pd} <2.5$ correlator $\langle \mathbf{s}\mathbf{S}_{i} \rangle=0  $ and
$E_{ex}$=0. For $J'_{pd}
> 2.5$ $E_{ex}$ increases  in accordance with the observed steps of $\langle
\mathbf{s}\mathbf{S}_{i} \rangle $. In Fig.~4b the total magnetic moment of the
polaron, $S_{iz}^{tot}= \sum_{i}S_{iz} + s_z$ is presented. It is also
characterized by  steps occurring for the same values of $J'_{pd}$ as those of
$\langle \mathbf{s}\mathbf{S}_{i} \rangle $ when $J'_{pd}$ is large. For
$J'_{pd} <9\,\,$ and 2D case the total magnetic moment is equal to $1/2$. For
$J'_{pd}<2.5\,\,$ this moment is located at the carrier,  for $J'_{pd}
>2.5\,\,$ it is a result of the coupling of the electron spin ($s_z=1/2$) with
$\sum_{i}S_{iz}=1$ AF spin. The correlator between two neighbor AF
spins $\langle \mathbf{S}_{i}\mathbf{S}_{i+1}\rangle $ is
presented in Fig.~4c. It changes from $-0.4$ for a pure AF to 0.25
(both spins parallel) with an increase in the strength of $p$-$d$
coupling. At the same time,  $S_{iz}^{tot}$ approaches value $7.5$
(all bonds broken). It is worth  to note that the step-like
behavior of $\langle \mathbf{S}_{i}\mathbf{S}_{i+1}\rangle  $ is
not a consequence of breaking of the subsequent bonds, but it is
associated with steps of the total magnetic moment of the system.

       To summarize, the ferron can form when homogeneous, Curie-Weiss
magnetization is induced. Our results complete well the classical
model of ferron.

\subsection{Comb-Like AFP}

A homogeneous distribution of the carrier on many AF spins makes the $E_{ex}$
on opposite spins cancel each other out. Thus we introduced a carrier trial
function which is distributed on one of the Neel sublattices only as
intuitively energetically favorable. The carrier distribution is described as
follows:

\begin{equation} \label{e22b}
|\varphi \left( i\right)|^2 =f\left( \mathbf{r}_{i}\right) \cos
^{2} \left( \frac{\pi }{2a} \mathbf{r}_{i}\right),
\end{equation}
where $f(\mathbf{r}_{i})$ is an envelope function, usually  Gaussian. We call
such a polaron   comb-like AFP \cite{HankiewiczPB,HankiewiczACT}.

$E_{ex}$ versus $J'_{pd}$ for the comb-like polaron with a Gaussian envelope
function is presented by triangles in Fig.1. For a small $r'_p$, the comb-like
AFP is equivalent to the small AFP. For a bigger  $r'_p$, the gain $E_{ex}$
decreases, but it is still much greater than that for  large AFP. $E_{ex}$ for a
large  comb-like AFP is only three times smaller than that of a small AFP. This
stems from  a decrease of quantum fluctuations (see part IIIF).

\begin{figure}[t]
\epsfxsize=.2in \centerline{\epsfig{file=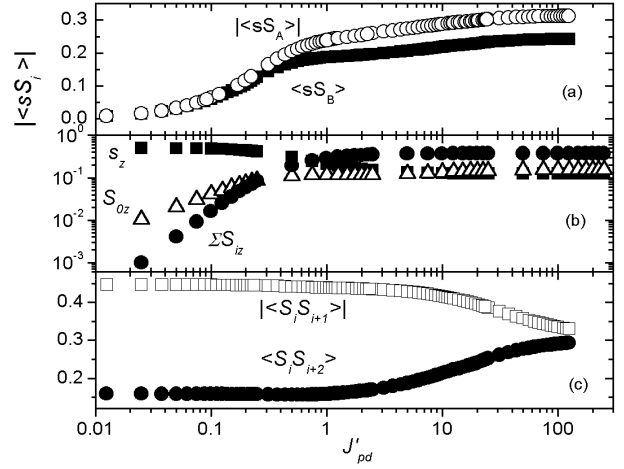,
width=8.cm,clip=}} \caption[]{(a) The mean value of the correlator
between the carrier spin $\mathbf{s}$ and the spins of the AF
sublattice on which the polaron is acting $\mathbf{S}_A$,
$|\langle \mathbf{s}\mathbf{S}_A \rangle|$ (open circles), and the
spins of other sublattice $\mathbf{S}_B$, $\langle
\mathbf{s}\mathbf{S}_B \rangle $ (filled squares) as a function of
$J'_{pd}$ for a comb-like trial function. (b) The z-component of
the carrier spin, $s_z$ (filled squares), the central spin
$S_{0z}$ (open triangles), as well as the sum of AF spins
 $\sum_{i}S_{iz}$ (filled circles) for an open chain of 16 spins as a function of
$J'_{pd}$. (c) The correlator between the spins from the same
sublattice $\langle \mathbf{S}_i\mathbf{S}_{i+2} \rangle$ (filled
circles) and from two neighbor sublattices $|\langle
\mathbf{S}_i\mathbf{S}_{i+1} \rangle |$ (open squares) as a
function of $J'_{pd}$. }
\end{figure}

The comb-like AFP induces a staggered magnetization in the AF medium. This is
shown in Fig.~5a, where the mean value of the correlator between the carrier
spin $\mathbf{s}$ and the spins of the AF sublattice on which the polaron is
acting $\mathbf{S}_A$, $|\langle \mathbf{s}\mathbf{S}_A \rangle|$, and the
spins of other sublattice $\mathbf{S}_B$, $\langle \mathbf{s}\mathbf{S}_B
\rangle $, are presented by circles and squares, respectively. The values of the
correlators increase linearly for small $p$-$d$ coupling, and they saturate when
$J'_{pd}>1$. The saturation value for $ \langle \mathbf{s}\mathbf{S}_B
\rangle=0.25$  and it is independent of the number of spins. The saturation
value of $|\langle \mathbf{s}\mathbf{S}_A \rangle|>0.25$ and depends
 on the number of
spins. The difference between the magnetization of the A and B
sublattices  gives the net magnetic moment of AF.

A part of the magnetic moment which is transferred from the carrier to the AF
medium depends on $J'_{pd}$, as it is shown in Fig.~5b. Here the change versus
$J'_{pd}$ of $z$- components of the carrier spin, $s_z$, of one of the spins on
which the carrier acts, $S_{0z}$, and  of the total  spin,
 $\sum_{i}S_{iz}  $, is presented.
  For small values of $p$-$d$ coupling a quadratic increase of
the moment transferred to AF  is observed.
 $\sum_{i}S_{iz}  $
 saturates at a value smaller than
0.5. Simultaneously, the z-component of the carrier spin decreases from 0.5 for
weak $p$-$d$ coupling to the saturation value of 0.15 for an AF cluster
consisting of 16 spins. However,
 $s_{z} +\sum_{i}S_{iz}=0.5  $
is independent on $J'_{pd}$.

The presence of the comb-like AFP on one of the Neel sublattices breaks the
translation symmetry and leads to the formation of two magnetized sublattices.
In Fig.~5c the correlators between NN, $|\langle \mathbf{S}_i\mathbf{S}_{i+1}
\rangle|$, and NNN, $\langle \mathbf{S}_i\mathbf{S}_{i+2} \rangle$, are
presented as a function of $J'_{pd}$. While $|\langle
\mathbf{S}_i\mathbf{S}_{i+1} \rangle|$ decreases for large $J'_{pd}$, the
correlator $\langle \mathbf{S}_i\mathbf{S}_{i+2} \rangle$ increases. Thus, a
comb-like AFP increases the AF correlation radius. In the limit of very large
$J'_{pd}$ and not too small sizes of the AF cluster, the comb-like AFP
transforms the quantum antiferromagnet into a classical one.

Although $E_{ex}$ for the comb-like AFP is large, the probability
of its formation is small because it is characterized by a  big
kinetic energy. Moreover, the comb-like AFP has a tendency to
localize into small AFP. This comes from the fact that the kinetic
energy is almost the same for both types of polarons (as long as
only $t_1$ is taken into account) while $E_{ex}$ due to quantum
fluctuations increases three times with localization.

\subsection{Quantum effects}

In this section we summarize the results concerning the formation mechanisms of
different magnetic polarons. We pay particular attention to the appearance of
quantum effects. In paragraphs A-E, five types of antiferromagnetic polarons
were presented. The formation of weak small, medium, and comb-like polarons
 is associated with the induction of staggered magnetization.
In contrast, the formation of a strong ZR polaron is caused by the
compensation of the AF spin by the carrier spin. The mechanism of
ferron formation is the breaking of AF bonds, which is equivalent
to the induction of a homogeneous  moment.

Classically, the only possible mechanism of AF polaron formation
at zero temperature is the turning of the spins in  the direction
of the effective field induced by the carrier spin
\cite{Kasuya68,Nagaev76}. However, this mechanism does not explain
the formation of the remaining polarons. The induction of
staggered magnetization is directly associated with the quantum
treatment of the AF.  In a quantum approach the ground state of AF
is a combination of Neel states, hence no sublattice is
distinguished. The simplest measure of AF quantum fluctuations is
the difference between the ground and the first excited states,
which are both composed of Neel states. In the presence of
staggered magnetization  the Neel sublattices are distinguishable
and the damping of spin fluctuations occurs \cite{Wilamowski00}.

The mechanism of compensation responsible for the formation of the
ZR polaron should be also considered in a quantum approach.
Classically, the carrier couples  the AF spin and such interaction
does not destroy the AF order. As we have shown in Fig.~3c, in the
quantum treatment of the antiferromagnet the affected AF spin may
cease to be correlated with neighbor spins. It means that only the
quantum approach allows to model the destruction of
antiferromagnetic order which is observed in experiments
\cite{Birgeneau88}. Despite the quantum character of the discussed
above mechanisms, there are some particular situations where  the
polarons can be described in the classical limit, but the quantum
corrections have to be taken into account.
 These corrections are the
result of scaling of spin fluctuations with the number of spins
and seem to be important for  the antiferromagnetic sign of
$J'_{pd}$. By a systematic analysis of the dependence of $E_{ex}$
on $p$-$d$ coupling in the range of very large $J'_{pd}$, we have
found that $E_{ex} =\sum_{i}j_{pd} (i)\langle \mathbf{s}
\mathbf{S}_{i} \rangle $, where the summation is taken over the
$N$ spins $\mathbf{S}_{i}$ within the polaron size (see Fig.~2).
The correlator depends on $N$ and is equal to $\langle
\mathbf{s}\mathbf{S}_{i}\rangle   = (0.25 +1/2N)$. Thus $E_{ex}$
changes from $0.75J'_{pd}$ for a polaron localized on one spin to
$0.25J'_{pd}$ for a polaron interacting with an infinite number of
spins. Hence, it is possible to calculate the energy of strong
small polarons as well as ferrons classically, but the appropriate
scalar spin multiplication (0.25) has to be replaced by  the
quantum factor. It approaches the classical
 value when $N\rightarrow\infty$,
whereas for $N=1$ it is three times bigger. The same quantum
factor 3 is lost when the ZR polaron is delocalized on one of the
Neel sublattices forming the comb-like polaron.

The importance of the sign of the $p$-$d$ coupling is shown in the
inset to Fig.~2.
 Here $E_{ex}$ for
a  small AFP with either a positive or negative  sign of $J'_{pd}$ is presented.
  Different slopes of $E_{ex}$ for
  ferro- and antiferromagnetic
  coupling are observed. It is a consequence of the fact
 that the carrier  and AF spins
  can not couple ferromagnetically to more than 0.25  while for antiferromagnetic coupling
   $\langle \mathbf{s}\mathbf{S}_i \rangle$ approaches $-0.75$.
  Thus, only for antiferromagnetic coupling the
  corrections associated with the number of spins in the AF system are important.
  In the case of ferromagnetic  coupling, the classical energy calculations
  for small polarons and ferrons are correct.

\subsection{Polaron mass}
In our approach (see Eq.(2)) we assume a simple decoupling of the
carrier kinetic energy and the spin degree of freedom.  For all
the  types of  polarons discussed, the minimum energy corresponds
to the localization of the maximum of the trial function,
$\Psi(\mathbf{r})$, exactly at the position of the local spin
$\mathbf{r_0}$. But another equivalent minima correspond to the
localization at another equivalent spin positions.  For PBC, which
simulate the translation symmetry, all minima are exactly
equivalent. As the consequence, the linear combination of the
simplified solutions is expected to be a better solution of the
problem. It shows that the carrier motion and the spin structure
are really coupled. To study that problem correctly one should,
however, diagonalize  $N$ times bigger matrixes. Anyway, our
simplified approach gives a reasonable picture of the spin
structure of a magnetic polaron under assumption of its static
nature, but it does not allow to study the real correlation of the
carrier motion and spin dynamics and does not bring any solid
answer on the problem of the polaron localization. However, a hint
about the effective polaron mass and the reduction of the kinetic
energy can be found within our simple approach through the
analysis of the change of the spin structure which accompanies the
transfer of the polaron to the NN. When the decoupling of the
motion and spin is assumed (Eq. (2)), the kinetic energy is
described by the parameter $t_1$, according to Eq. (6). Here, only
the transfer without any spin flip is considered.  Our solutions
show, however, that the transfer is accompanied with a complex
change of the spin structure. The spin structure of the
surrounding local spins is very different when the carrier is
localized on the first, A, or on the neighboring site, B. A
posteriori, we are able to claim that the matrix elements
describing the kinetic energy are reduced by a factor resulting
from the  product of the spin states on the site A and B $\langle
X_S^A |X_S^B \rangle$, respectively. In other words, to consider
the effects which are omitted due to our assumption (Eq.(2)), one
has to reduce the value of the parameter of the kinetic energy.
Instead of $t_1$  as effective kinetic energy, $t_1^*= t_1\langle
X_S^A |X_S^B \rangle$ should be used. Our numerical study shows
that for weak $p$-$d$ coupling, $J_{pd}\ll J_{dd}$, the factor
$\langle X_S^A |X_S^B \rangle$ is close to unity. For
$J_{pd}\approx J_{dd}$ the value of $\langle X_S^A |X_S^B \rangle$
decreases, and for $J_{pd}> J_{dd}$ it saturates at the value in
the range 0.1 - 0.2, depending on the type of polaron. It shows
that the effective mass of polaron dressed by the spin
polarization is considerably bigger as compared to free polaron.

\section{PHASE DIAGRAM OF MAGNETIC POLARONS}

Until now we have considered the $E_{ex}$ from magnetic polarons formation and
we have described the spin structure of various polarons. However, the total energy
(exchange, phonon and kinetic one) should be taken into account to specify
which type of polaron forms. In this section we consider the formation of a
pure magnetic polaron.

\begin{figure}[tbh]
\mbox{}\centerline{\epsfig{file=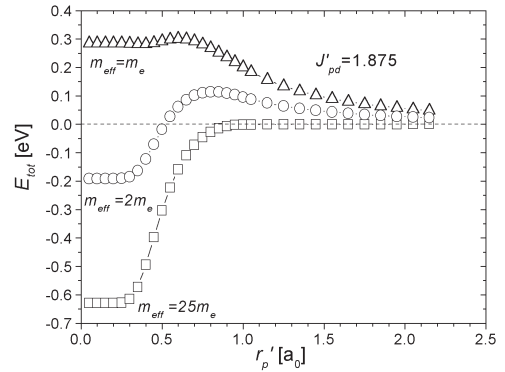,width=6.5cm}}
\caption[]{Total energy, $E_{tot}$, as a function of polaron
radius $r'_p$ for $J'_{pd} =1.875$ and effective masses of $m_e$
(triangles), $2m_e$ (circles), $25m_e$ (squares).}
\end{figure}

In Fig.~6 the total energy (sum of kinetic and exchange energies) as a function
of polaron radius calculated for a few chosen effective masses is shown for a
2D square cluster with PBC. It can be seen that for $m_{eff}=m_e$ no AFP can
form. For a large $m_{eff}$ (which scales as $1/t_1$), the only polaron which
can exist is the small one. The formation of a ferron is not possible because
$J'_{pd}$ is too small. The medium and large polarons tend to localize in one
elementary cell, independently of the values of $m_{eff}$ and $t_1$. The
tendency to localization is associated with a rapid reduction of $E_{ex}$ with
 increasing $r'_p$, which stems from
quantum fluctuations of the AF medium (see part IIIF). We  found that for
$J'_{pd} =1.875$ magnetic polarons can form only if $m_{eff}>1.6m_e$, what
corresponds to $t_{1}=2J_{dd}$.
  For weaker $p$-$d$ couplings the small AFP is
also the only polaron which can form but its $m_{eff}$ should be  much greater.

In Fig.~7 the phase diagram of the formation of different magnetic
polarons is shown.
  The area of parameters $t_1$ and $J'_{pd}$ for which the small
   AFP forms is marked  by a solid line.

\begin{figure}[tbh]
\mbox{}\centerline{\epsfig{file=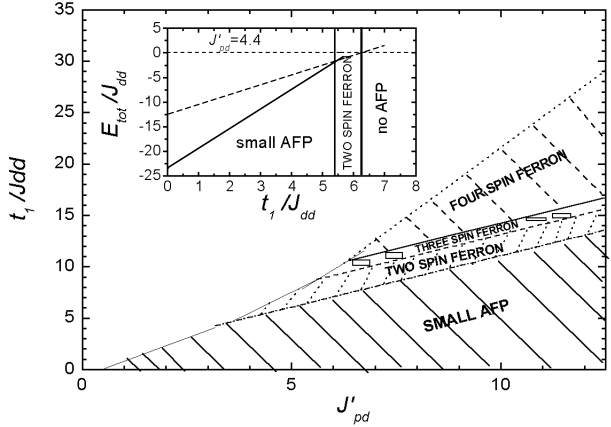,width=8cm,clip=}}
\caption[]{Phase diagram of  magnetic polaron formation.  In the
inset  $E_{tot}$ as a function of $t_1$
 for $J'_{pd}= 4.4$ is shown. In the inset dashed line presents $E_{tot}(t_1)$
 for two-spin ferron, the solid line shows $E_{tot}(t_1)$ for small AFP. Vertical lines separate
 the areas of different polarons formation}
\end{figure}

For higher $J'_{pd}$,  the small ferrons can also form. In the inset to Fig.~7
the total energies of a small AFP and a small ferron are compared for
$J'_{pd}=4.4$.
  $E_{ex}$ for a small AFP is higher than
for the two spin ferron, and it is constant. The kinetic energy of
two-spin ferron (polaron localized on two neighboring spins) is
smaller than that of a small AFP  and scales linearly with $t_1$.
Thus the total energies of both polarons are linear functions of
$t_1$ but with different slopes. In particular the boundary
between small AFP and two spin ferron for $J'_{pd}=4.4$ crosses
for $t_1= 5.4J_{dd}$, what is equivalent to $m_{eff}\geq 0.6m_e$.
For $t_1$, which is greater than $6.25J_{dd}$ any AFP is not
formed. In general, the small AFP forms for $m_{eff}\geq m_e$,
while the two-spin ferron appears for $m_{eff}<m_e$.

 Similarly, we found the range of parameters for which
  other types of polarons can also form.
In Fig.~7 the ranges where  small AFP, two, four-spin ferrons
  can occur are shaded
by solid, dotted and dashed lines, respectively. The appearance of a three-spin
ferron is indicated by rectangles. Larger ferrons appear for much stronger
$p$-$d$ couplings, and we omit their description because they would exist for
nonphysical parameters.

The presented results were calculated for a 2D square with PBC. To
estimate the error caused by small size of antiferromagnetic
cluster, we investigated also such clusters as: a  $4\times4$
square without PBC, a $4\times4$ parallelogram, and $2\times N$
systems where $2\leq N\leq10$. We found that the energy of
polarons with small size depends on the nearest vicinity  only. It
is weakly dependent on both $N$ and  boundary conditions. For
different 2D clusters the formation range of subsequent polarons
can be shifted about 10\%--15\%.

The investigation of polaron formation in a wide range of parameters is
exceptionally valuable when we want to classify what type of polaron could be
formed in different materials. For example, in doped EuTe  where $J_{pd}\approx
0.125$~eV and $J_{dd}\approx2$~meV \cite{Wachter79} the $p$-$d$ coupling on one
AF bonds $J'_{pd}\approx 8$. At the same time, $t_1\approx15J_{dd}$. For such
parameters our model predicts the ferrons formation (see Fig.~7) in accord with
experiments (see \cite{Kasuya68,Wachter79} and references therein).

In high-$T_c$ superconductors which are based on CuO$_2$ layers, the $d$-$d$
exchange coupling ranges from 0.055~eV to 0.085~eV, as determined from the Neel
temperature of antiferromagnetic precursors of these materials\cite{Plakida}.
  $J_{pd}\approx 1-3$~eV was determined  from the  band
structure calculations\cite{Buhler00,Hybertsen00,Eskes00,CXChen90}. Thus, the
ratio of $p$-$d$ to $d$-$d$ exchange energies is between 1.5 and 4 for
$J_{dd}=0.075$~eV. $m_{eff}$ is much more difficult to determine because the
mass observed in experiments (e.g. in ARPES) is decorated by phonon and
exchange interactions. Since the decorated $m_{eff}$  observed in ARPES
experiments \cite{ManteFink90} is about $4m_e$ and $m_{eff}$ in our model is
dressed only by phonons, the $m_{eff}$ which  should be considered in the phase
diagram is smaller than $4m_e$, which is equivalent to $t_{1}
> 0.8J_{dd}$.

We conclude on the basis of our phase diagram that for parameters typical for
weakly doped high-$T_c$ superconductors mainly the small AFP would  form. There
may be also some competition between the formation of small AFP and two-spin
ferrons provided that the effective mass dressed in phonons is a bit lower than
the electron mass.  Large AFP have  energies of few orders of magnitude lower
than those of small AFP and ferrons.

\section{THE CONTRIBUTION OF PHONON AND MAGNETIC POLARON}

\begin{figure}[tbh]
\mbox{}\centerline{\epsfig{file=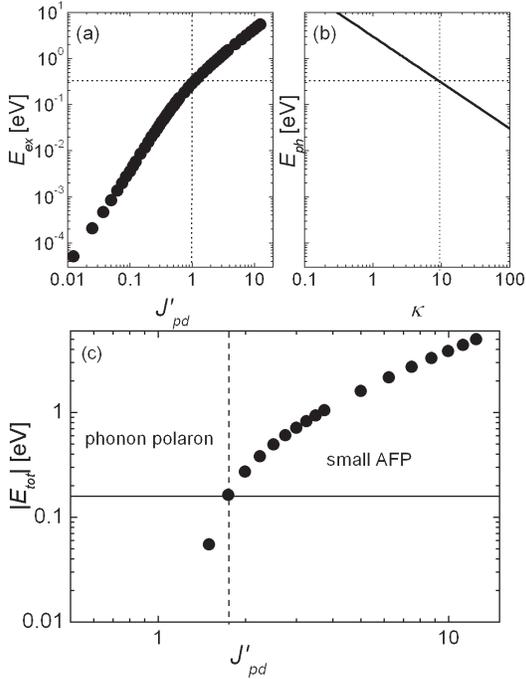,width=7.0cm,clip=}}
\caption[]{$E_{ex}$ from the formation of a small AFP as a function of
$J'_{pd}$(a). The phonon energy, $E_{ph}$, versus effective dielectric constant
$\kappa$ (b). The absolute value of the total energy for a magnetic polaron
(circles) and for a phonon polaron (solid line) as a function of $J'_{pd}$ (c).
Here, $m_{eff} =2.2 m_e$ and $\kappa = 5$. The vertical dashed line shows the
value of $J'_{pd}$ where the magnetic and phonon polaron energies are equal.}
\end{figure}

Now we will compare the phonon and magnetic contributions to the
formation of polarons in weakly doped high-$T_c$ superconductors.
We have shown that in the 2D case $E_{ex}$ from the formation of
large magnetic polarons is close to zero and that they are
unlikely to form (Fig.~6 and Fig.~7).  Mott and Alexandrov
\cite{Alexandrov95} suggest that in high-$T_c$ superconductors
only small phonon polarons form (see Eq.~(12)). Thus we will  not
consider  large polarons here, neither of magnetic nor phonon
origin.

In Fig.~8a, $E_{ex}$ due to the formation of a small magnetic
polaron in a 2D AF square  with PBC is shown as a function of
$J'_{pd}$. In Fig.~8b, the energy gain due to the formation of
phonon polaron is presented as a function of effective dielectric
constant $\kappa$. Typical values of $\kappa$ are between 5 and
20. The kinetic energy for both small polarons is the same,
 $E_{k} =4\left| t_{1} \right| $,
and is omitted  for clarity.

Horizontal and vertical dotted lines  drawn in Figs. 8a and 8b indicate
parameters $J'_{pd}=1$ and  ${\kappa}=10$ for which energy gain
$E_{ex}=E_{ph}=0.35$~eV is the same for both phonon and magnetic polarons.
The energy gain due to the formation of phonon polarons is greater than that
for magnetic polarons for ${\kappa}<10$ and $J'_{pd}< 1$, while ${\kappa}>10$
and $J'_{pd}>1$ the magnetic polaron dominates. One can see that for the range
of parameters, which seem to be typical for high-$T_c$ superconductors, the
energy gain due to the formation of phonon and magnetic polarons is comparable
and  of the order of fraction of eV.

 In order to obtain the total energy  of magnetic and phonon
polaron separately, $E_{tot}$, we have evaluated the energies for a chosen set
of parameters: $m_{eff} =2.2m_e$ and $\kappa = 5$. This mass, according to
ARPES experiments\cite{ManteFink90} and band structure calculations
\cite{Hybertsen00}, is reasonable for high-Tc superconductors.  $\kappa$ is
taken from theory of phonon polaron \cite{Alexandrov95}. In Fig.~8c, we show
the values of $E_{tot}$ for phonon as well as magnetic polarons versus
$J'_{pd}$. For the phonon polaron $E_{tot}$ is constant and equal to 0.15~eV
since
 the phonon as well as kinetic energies are
independent of $J'_{pd}$.
 The phonon polaron dominates in the
range of weak $p$-$d$ couplings.  In contrast, the magnetic polaron
(circles in Fig.8c) dominates in the range of strong $J_{pd}'$
where its energy increases linearly with $J_{pd}'$. In the range
of $J'_{pd}= 1.5 - 4 $, which is typical for weakly doped
high-$T_c$ superconductors, the energies of both polarons are
comparable or magnetic polaron energy slightly overcomes the
energy of phonon one. This result agrees well with a very recent
paper \cite{Lanzara01} which stresses the importance of phonon
interactions in high-$T_c$ superconductors.

We conclude that within our model the  phonon and magnetic polaron energies are comparable and that both  contributions
should be included in the total polaron energy  in high-$T_c$ superconductors.
\section {SUMMARY}
Using the spin-fermion model and treating all spins in quantum
approach,  we found that depending on the material parameters
 five various types of magnetic polarons in antiferromagnetic medium
 can be distinguished.

We showed that in the range of parameters typical for weakly doped high-$T_c$ superconductors the contributions of
phonon and magnetic interactions to the formation of a polaron are comparable and hybrid magnetic-phonon polarons
form.

 In  weakly doped high-$T_c$ superconductors small magnetic polarons as well as small ferrons can form. Our
numerical study allows to find a continuous evolution from Zhang-Rice approach, when $p$-$d$ coupling is assumed
to be much stronger than $d$-$d$ coupling, via important for high-$T_c$ materials intermediate range, to linear
response approach suitable for weak  $p$-$d$ coupling.

 The here introduced  comb-like polarons cannot form when the band is empty (i.e. in
undoped material) because of the cost of high kinetic energy. However, they can
play a fundamental role in doped high-$T_c$ materials when the band is
partially filled \cite{HankiewiczPB,HankiewiczACT}.\section*{ACKNOWLEDGEMENTS}
Valuable discussion with H. Przybyli\'nska. Work supported by the KBN grant 2
P03B 007 16.\hfill
\bibliography{han2201}

\end{document}